\documentclass[prl,aps,twocolumn]{revtex4}
\usepackage{graphicx}
\begin{document}
\title{Bulk ferromagnetism and large changes in photoluminescence intensity  by magnetic field in $\beta$-Ga$_2$O$_3$. }
\author{ V. Sridharan $^1$, Sangam Banerjee $^2$, Manas Sardar $^1$,  Sandip Dhara $^1$, 
N. Gayathri$^{1,2}$ and V. S. Sastry $^1$}
\address{$^1$ Material Science Division, Indira Gandhi Center for Atomic Physics, kalpakkam 603102, India}
\address{$^2$ Surface Physics division, Saha Institute of Nuclear Physics, 1/AF Bidhannagar, Kolkata, India}

%\maketitle
 
 \begin{abstract}

   In this letter we report observation of room temperature ferromagnetism in bulk
Ga$_2$O$_3$ . We also observe large (10-60$\%$) increase/decrease in photoluminescence
In the red(700 nm wavelength)/blue(500 nm), with the application of small magnetic field(0.4 Tesla). We argue, 
that ferromagnetism occurs entirely due to  chains of oxygen(O(3) sites, Fig. 5) vacancies.
We propose a simple model to explain, origin and location  of moment, formation of ferromagnetic dislocation needles,
and strong increase/decrease ofred/blue photoluminescence intensity with magnetic field.

\end{abstract}
 

\maketitle
%\begin{multicols}{2}

Ever since the discovery of ferromagnetism in graphite\cite{graph}, CaB$_6$\cite{bor}, HfO$_2$\cite{HfO2} a 
new paradigm of ferromagnetism has opened up, which Coey\cite{d0-ferro} aptly calls d$^0$ ferromagnetism.
Unlike traditional ferromagnets where moments come from unfilled
d or f electrons, these materials have no obvious source of magnetic moment.
 Recently ferromagnetism was also reported in TiO$_2$ and In$_2$O$_3$ thin films\cite{thinfilm}.
 Ferromagnetism in nanoparticles of Al$2$O$_3$, ZnO, SnO$_2$ , CeO$_2$ is observed by Sundaresan et al\cite{Sundaresan} and 
conjectured to be a universal property of metal oxides with either empty or filled d or f energy levels,
 in nanoparticle form. There are strong phenomenological reasons to believe that the common origin of ferromagnetism in all 
these materials might be due to defects and vacancies in the solid\cite{d0-ferro}.
 Theoretically Monnier et al\cite{boride-theo} have predicted moment formation near missing B$_6$ octahedra(anion vacancy)in 
CaB$_6$. Elfimov et. al \cite{sawatzky} had predicted small moment formation near cation vacancy in Rocksalt structure
 binary metal oxides(CaO). The basic idea is to have atomic vacancies that have large local symmetry, 
so that electrons/holes residing in the vacant sites have degenerate orbitals available.
 Local Hund's exchange forces parallel alignment of the defect site electrons and give rise to non zero moment.
 Experimentally we dont know much about the spin structure in these materials and the role of different kind of defects
 are also not known. Neither do we have a chemical intuition about the requisite kind of defects that might hold moments. 
In this paper, we report bulk(grain size larger than 30 micron) ferromagnetism in Ga$_2$O$_3$.
 To our knowledge nobody has found this before. We also find a large(15-20 $\%$ increase in photoluminescence 
intensity with application of  small magnetic Field(0.4 Tesla). Starting from the structure of Ga$_2$O$_3$ we present
 a simple intuitive model , that identifies O(3) type anion vacancies(see Fig. 1) as the one that shows moment. 
Our model helps in understanding, ferromagnetism as well as increase in Photoluminescence with magnetic field in a unified way. 

%Ferromagnetism and photoluminoscence intensity rise with magnetic field, occurs due to the defect site(anion vacancy) 
%electrons with energy levels within the Band gap. The valence and conduction band electrons are not directly involved.
% Yet the crystal structure of template wide gap semiconductor is important, so that one have the required kinds of defect sites. 
%In Ga$_2$O$_3$ we propose a mechanism for ferromagnetism and PL intensity change with applied magnetic field.

Ga$_2$O$_3$ powder of $99.999\%$ pure was obtained from Alfa Aesar, UK. Powder was calcined at $950^o$C in static air for $24$ hours and 
furnace cooled.  Powder was compacted into pellets of $10$ mm diameter under a pressure of $70$ MPa.
 The pellets were annealed for $24$ hours at $1200$ and $1300^o$C in a closed environment of low oxygen fugacity.
 Room temperature X-ray powder diffraction (XRD) pattern was recorded in 2q range $15$ to $120$ degrees with a 
step interval of $0.05$ degrees using STOE diffractometer operated in Bragg-Brentano geometry.
 Magnetization (M) as a function of temperature and field (H) measurements were carried out in a MPMS-7 Tesla
 Quantum Design system from $2$ K to $340$ K. Optical photoluminescence (PL) was studied using $324$ nm line of 
He-Gd continuous wave laser as excitation and dispersion with three set of holographic blazed $1800$ grooves/mm gratings
 in a double subtractive triple monochromator (Juvin-Yvon T64000 spectrograph) in a back scattering configuration.
 A liquid N2 cooled back-thinned CCD detector is used for the detection of scattered intensity. 
The room temperature (RT) powder X-ray diffraction of $\beta$-Ga2O3 is shown in Figure 1 and the inset of Fig. 1 shows the 
Raman spectra recorded at RT. The diffraction pattern could be indexed to monoclinic structure (Space group: C 2/m).
 The zero field cooled (ZFC) and field cooled (FC) thermomagnetisation curves for H=100 and 200 Oe are shown in Fig. 2.
 It is seen that the magnetization for H= 200 Oe larger that for H=100 Oe and that they exhibit a temperature of irreversibility ($T_{irr}$)
 at about 300 K, clearly indicating the presence of ferromagnetic ordering. In all the measurements, 
a dominant paramagnetic contribution emerges for $T < 20$ K. The DM (MFC-MZFC) vs T plot (inset of Fig 2) demonstrates 
the presence of a robust ferromagnetic state with a Curie transition temperature $T_c = 303$ K. 
The hysteresis loops recorded at T= 150, 55 and 2K are shown in Fig. 3. The loops exhibit vanishingly 
small remanent magnetization ( $12\times 10^{-3}$ emu/g) and coercive field ( 136 Oe) (Fig. 3(a)), much similar to that of
 diluted magnetic semiconductors. Though the loops exhibit a closure, they are progressively sloppy with decrease of
 temperature signifying the presence of paramagnetic contribution, consistent with thermomagnetisation results (Fig. 2).

\begin{figure}
\includegraphics[width=8.5cm,height=5cm]{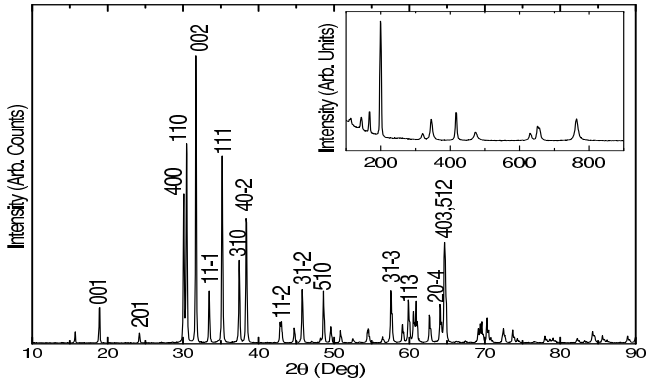}
%\vskip 4cm
\caption{ \label{Fig:1} X-Ray power diffraction pattern . Inset shows the Raman spectra.}
\end{figure}

\begin{figure}
\includegraphics[width=8.5cm,height=6cm]{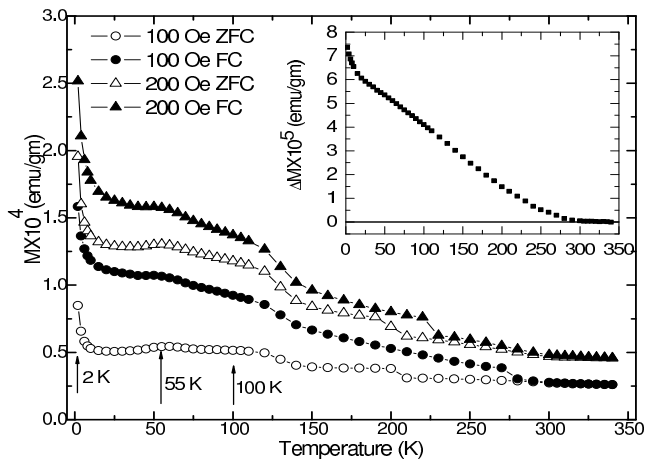}
\caption{\label{Fig:2} Field cooled and zero field cooled magnetization
versus temperature at fields of 100 and 200 Oe. Inset shows DM (MFC-MZFC) vs T plot.}
\end{figure}

\begin{figure}
\includegraphics[width=8.5cm,height=6cm]{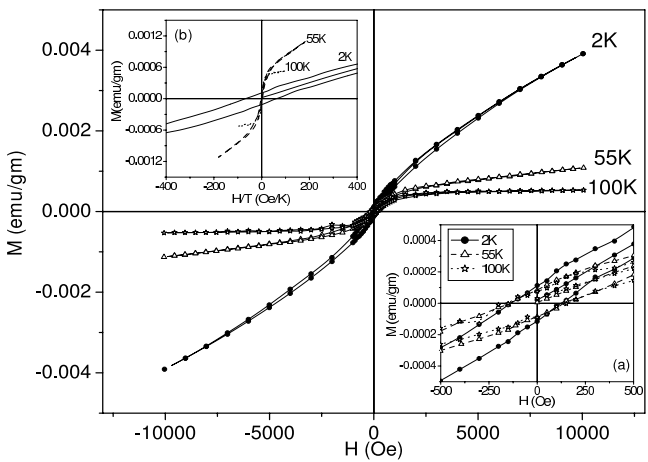}
\caption{\label{Fig:3} Hysteretic M-H loops at T=150, 55, 2 K. Inset Fig 3(a) expands 
the low field region, and Fig 3(b) shows M vs ${H\over T}$ plot showing that 
what we see is ferromagnetism and not superparamagnetism.}
\end{figure}

 The photoluminescence measured at RT (see Fig 4) exhibits a predominant red(725 nm) and weaker UV(395 nm) and blue 
(525 nm) emission. UV, blue PL were found to be dominant emissions in single crystals\cite{harwig}, they are substantially
 suppressed in the ceramic samples\cite{harwig} . 
Striation riding over the red emission (700 to 900 nm) is
 artifact of the back thinned CCD camera and is known to arise from the Ettalong effect.

\begin{figure}
\includegraphics[width=8.5cm]{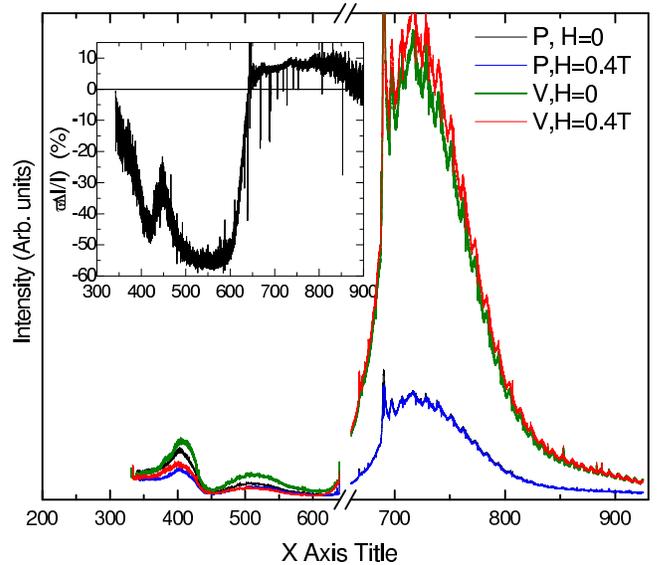}
\caption{\label{Fig:4} Photoluminescence(emission) spectra without(green line) and with(red line) external magnetic field( 0.4 Tesla)
for $\beta$- Ga$_2$O$_3$ vacuum annealed at 1000 K for 24 Hours. Blue line is for Prestine sample. Inset shows the percentage change
in intensity with magnetic field.}
\end{figure}
         Room temperature electron spin resonance (ESR) studies 
carried out in X-band mode using Bruker instrument could 
not detect any paramagnetic species. From the particle induced X-ray emission (PIXE) 
studies under the bombardment of proton, no impurity species could be detected except 
for Ni, whose concentration is determined to be less than 200 ppm. From 
the ESR and PIXE measurements it is clear that, ferromagnetism in Ga$_2$O$_3$ does
 not arise from the paramagnetic impurity species. Though, non saturation of 
hysteresis loops at 55 and 2 K (Fig. 3) could imply superparamagnetism, it is 
to be noted that the same measured at higher temperature (100 K) exhibits 
a saturation and no blocking temperature, as seen in 
the thermomagnetisation curves (Fig. 2). Additionally, the nonscaling of M vs H/T 
curves (Fig. 3(a)) rules out superparamagnetism. Strong deviation of the loops
 from saturation with decrease of temperature signifying the presence of paramagnetic contribution, whose origin will be discussed below and is consistent with thermomagnetisation
results (Fig. 2).  From the foregone discussion, we conclude that a long range ferromagnetic 
ordering is present. 
%and ascribed to due to defects, in similar to 
%that of other d0 ferromagnetic materials. 
In the following a plausible explanation 
is provided for presence of ferromagnetic interaction, intimately related to defect structures. 

        Substantial difference exists, atleast in a quantitative way, 
in the defect structures between poly and single crystals. Harwig et al
 report emission of UV and blue/green PL signals from single crystals. 
%even with the lowest interaband excitation energy, provided that $T > 120$ K\cite{harwig} . 
These PL signals are related to presence of acceptor states(see Fig 7) 
present in the crystals due to evaporation 
of GaO$_2$ during the growth process. Such defect states 
are less in pure polycrystalline samples and the 
corresponding(blue) PL signals are suppressed (Fig. 4). This also explains, 
in part, the absence of ESR signal in our sample  
, which are otherwise present in the single 
crystal samples\cite{aubay}.

Ga$_2$O$_3$ is a ionic insulating dielectric with band gap of  4.7 eV\cite{tuppins}.
 Eletrical conductivity is entirely due to defect electronic States\cite{many1,many2} with n type conductivity and 
arises due to oxygen vacancies. Concentration of charge carrier estimated by Hall measurement{many2} is about $10^{18}/$cm$^3$ 
to $3.0 \times 10^{17}$ cm$^3$ for single crystal and polycrystalline materials.
Several experiments point out partial delocalisation of electrons in Ga$_2$O$_3$ , even though conductivity is
 always activated type with small gap(20-30 meV). There are dynamic nuclear magnetic polarisation by Overhauser effect\cite{Overhauser} 
and narrow EPR lines(coming from motional narrowing)\cite{ESR-line} which points to presence of mobile electrons. 
PL measurements\cite{Blue-lumino} on the other hands shows large Stoke shift,
 characteristics of localised electrons with large electron-phonon coupling. 
Conductivity is highly anisotropic in single crystals. Conductivity along b axis is an order of magnitude larger than in other directions\cite{Ueda}

\begin{figure}
\includegraphics[width=8.5cm,height=7cm]{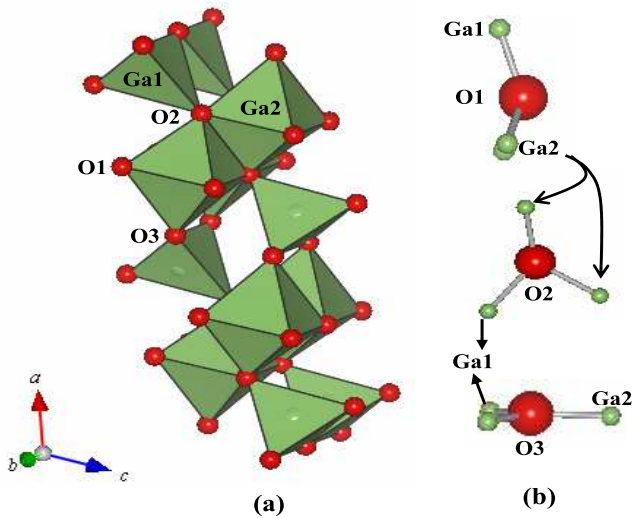}
\caption{\label{Fig:5} Structure of $\beta$-Ga$_2$O$_3$ showing three different types of Oxygen sites O(1),O(2) and O(3). The local symmetry around these sites are shown at the bottom.}
\end{figure}

\begin{figure}
\includegraphics[width=7cm,height=2.5cm]{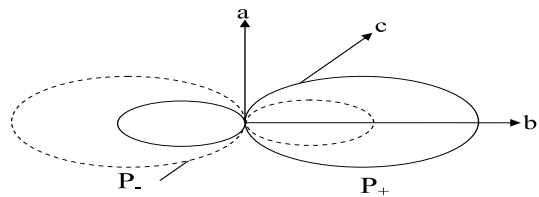}
\caption{\label{Fig:6} Schematic drawing of doubly degenerate donor electron wavefunctions $P_{\pm}$ around a O(3) vacancy. }
\end{figure}

\begin{figure}
\includegraphics[width=8.5cm,height=4cm]{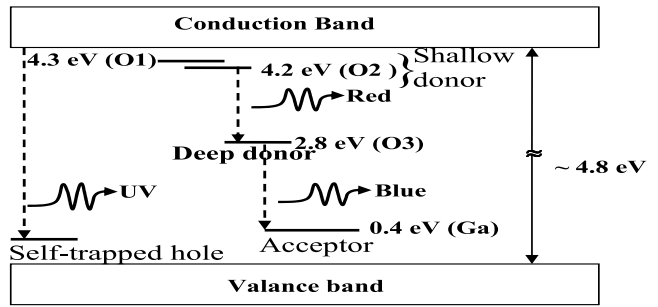}
\caption{\label{Fig:7} Schematic drawing of defect site electron enrgy levels
and the electronic transions responsible for UV, Blue and Red emission
in $\beta$-Ga$_2$O$_3$.}
\end{figure}

   Oxygen O(1) sites are not in plane of 3 near neighbour Ga atoms. O(2) sites has tetrahedral coordination with 4 Ga
atoms.
O(3) sites are at the center 
of the plane made by 3 near neighbour Ga atoms , and forms 
a chain along the b axis(perpendicular to the plane of 3 Ga atoms). Local site symmetry of O(3) sites are much more than
O(1)/O(2) sites.
The lowest energy wavefunctions of the O(3) vacancy site electrons will
presumably be like, $P_{\pm}={1\over \sqrt 2}(S\pm P_z)$ , where z is along the 
chain b axis. This symmetry related degeneracy , forces the two electrons 
on the neutral O(3) vacancy sites to occupy singly two degenerate
$P_{\pm}$ orbitals in a spin triplet $S=1$ state. $P_{\pm}$ wavefunctions(Fig 6) have considerable over
overlap(large hopping matrix element) along the chain direction.
To have double exchange operating between the O(3) vacancy sites
we propose the existence of of some fraction of O(3) vacancy sites 
with one electron
%($F^{+}$ center, i.e $O_V^'$ in Kruger-Vink notation),
 with the other eletron
 promoted to a nearby Ga atom which moves to an intertitial position. 
This is a O(3) Vacancy-Ga intertitial pair(PR from now on). This pair has a vacancy in either $P_+$ or $P_-$ orbital around the O vacancy site. This vacancy is a carrier.
So if the O(3) vacancies are bunched along chain segments, each having some number of PR's ( charge carriers) then, local  ferromagnetism follows naturally. The double 
exchange and electronic delocalisation is happening only on the O(3) vacancy chains . 
The original $Ga_2O_3$ lattice sites acts as just a template. Isolated O(3) vacancies
will give paramagnetic signals. O(3) vacancy chains without
sufficient number of $F^+$(PR) centers are likely to be antiferromagnetic(superexchange) and not
contribute to total moment. Assuming 0.2$\%$ of O(3) vacancies (each holding about 1 $\mu_B$ moment)
and assuming about half of them taking part on double exchange along fragmented chains, we estimate a moment of $5\times
10^{-3} $ emu/gm , which is close to the observed monemt($8\times 10^{-3}$ emu/gm) at H=7 Tesla and T=2 K.

Large $T_c$, low value of the net moment, the persistence of some paramagnetic component to
the total moment at the lowest temperatures, all  makes the bulk Ferromagnetism problem very challenging.

  Photoluminoscence comes from transition of electrons between different kinds of defect
related(centered) states. Their energies all fall within the Band gap.
Hajnal et al has estimated by semiemperical quantum chemical calculations, the energy levels of donor
electrons coming from O(1),O(2) and O(3) sites at 4.2,4.3(shallow donor) and 2.8(deep donor) eV.
In Fig. 7 we schematically draw different levels and show the photoluminoscent transitions.
We believe the red emission at 700 nm( 1.3 to 1.7 eV) is predominantly due
to electronic transitions when, a donor electron from O(1) or O(2) vacancy site 
jumps to the unoccupied ($P_+$ or $P_-$) level of a PR.
 The electron that jumps down to the O(3) sites must come with right
spin direction(parallel to the spin of the already existing electron .
This local Hunds rule constrain affects the transition in the red region.
The net spectral weight under the red emission line, W is given by,
$$
W \propto N_1 \cdot N_2 \cdot N_p \cdot < \cos({\theta_{1,2}-\theta_3 \over 2}) >
$$
where $N_1$ and $N_2$ are the number of electrons in O(1) and O(2) vacancies, $N_p$ is the number of
O(3) vacancy-Ga intertial pair, The cosine factor
depends on the relative spin  orientations
of the electron in the initial state at O(1) or O(2) site and the lone occupied electron 
present in the final state O(3) vacancy site. The incoming electron must come with correct spin.
This is the origin of the strong increase of the Red emission net spectral weight
with the application of very small  magnetic field. For Blue emission on the other hand , the final state
is an acceptor state which is nondegenarate. So if it already contain an electron , it can only accept
an electron of opposite spin. This becomes difficult with application of magnetic field. That is why Blue/UV
emission intensity gets suppresed with magnetic field.

To summarise, we found Bulk ferromagnetism in $\beta$-Ga$_2$O$_3$ . We also find large increase/decrease in 
photoluminescence in the red/blue region, with application of small magnetic field(0.2 Tesla). We propose, the magnetic moments
arise from O(3) vacancy chains(dislocation lines). Recent works on Fe doped HfO$_2$\cite{vacancy1}
and Co doped TiO$_2$\cite{vacancy2} points out that magnetism in these 
dilute magetic semiconductors could be due entirely to the anion
vacancies. Ubiquitous presence of charged dislocations in ZnO\cite{dis-charge} 
and GaN\cite{dis-charge,dis-cond-mobi} improving conductivity(magnetism ?) all suggests that
magnetism in these materials may be improved by controling density of dislocations. Light induced magnetism
and large chage in photoluminescence intensity with magnetism in all these materials should be looked for.

        We thank J. Subramanian and  Victor Babu, Centre for Leather
 Research Institute, Chennai for the ESR measurements and B. Sundaravel, IGCAR for the PIXE
measurements, and G. Ragavan, Sharat Chandra and B. Panigrahi, IGCAR for discussions.

%\end{multicols}{2}
\end{document}